\documentclass[aip,rsi,reprint,groupedaddress]{revtex4-1}
\usepackage{graphicx}  

\begin{document}


\title{
Integrated Electronic Transport and Thermometry at milliKelvin Temperatures and in Strong Magnetic Fields
}

\author{N. Samkharadze$^1$, A. Kumar $^1$, M.J. Manfra$^{1,2}$ , L.N. Pfeiffer$^3$, K.W. West$^3$, 
        and G.A. Cs\'{a}thy$^1$ \footnote[1]{gcsathy@purdue.edu}}

\affiliation{${}^1$ Department of Physics, Purdue University, West Lafayette, IN 47907, USA \\
${}^2$ Birck Nanotechnology Center, School of Materials Engineering, and School of Electrical and Computer Engineering,
Purdue University, West Lafayette, IN 47907, USA \\
${}^3$Department of Electrical Engineering, Princeton University, Princeton, NJ 08544\\} 


\date{\today}

\begin{abstract}

We fabricated a He-3 immersion cell for transport measurements of semiconductor nanostructures 
at ultra low temperatures and in strong magnetic fields. We have a new scheme of field-independent
thermometry based on quartz tuning fork Helium-3 viscometry
which monitors the local temperature of the sample's environment in real time. 
The operation and measurement circuitry of the
quartz viscometer is described in detail. We provide evidence
that the temperature of two-dimensional electron gas confined to a GaAs quantum well follows the
temperature of the quartz viscometer down to 4mK.

\end{abstract}
\maketitle

\section{Introduction} 

One of the most exciting subjects in contemporary condensed matter physics is the study of the emergent phenomena in correlated electron systems. In particular, the two-dimensional electron gas has a very rich physics and
new phenomena are expected to occur as the temperature is lowered.
In the absence of a magnetic field a transition to a ferromagnetic state 
is predicted with decreasing electron density \cite{stoner1964,candido2004}. In strong magnetic fields 
recent theories predict that certain fractional quantum Hall states could have
exotic excitations obeying non-Abelian statistics \cite{moore91}. Extending transport experiments to ultra low temperatures 
is expected therefore to lead to the discovery of new electronic ground states \cite{pan02,eisen02,xia04,kumar2010} and offers the rare chance of finding a new class of non-Abelian particles \cite{willett09}.
Such work will also contribute to a better understanding of the phenomena in closely related strongly correlated systems such as the exotic p-type superconductivity in strontium ruthenate, p-type superfluidity in He-3, and 
condensates in atomic gases of the Fermi type \cite{stern10}.

A critical experimental capability for studying correlated ground states is the achievement of ultra
low electronic temperatures in transport measurements of systems such as GaAs.
While modern dilution refrigerators can routinely cool below 10mK,
such low electronic temperatures in semiconductor nanostructures are often difficult to achieve. There are several mechanisms which limit cooling of the electrons. First, cooling of a semiconductor crystal is limited by the Kapitza thermal resistance due to the phonon mismatch of the semiconductor and its thermal environment \cite{pobell}. Second, 
electrons couple to the host lattice via the electron-phonon coupling which becomes extremely weak at low temperatures \cite{wennberg1986}. Cooling of the electrons  occurs therefore mainly through the measurement leads heatsunk to the cold spot of the refrigerator.
Finally, even with the most careful electromagnetic filtering there are minute quantities of energy traveling on the electrical leads in form of high frequency waves. Because of the poor thermal contact of the electrons described above these waves will often heat electrons above the minimum temperature of the refrigerator.

We describe a new design for a He-3 immersion cell which allows cooling and transport measurements of electrons in  semiconductor nanostructures down to 5mK and which allows for a simultaneous real time monitoring of 
the temperature at the location of the sample.
Temperature is measured using a quartz tuning fork which monitors the viscosity of the He-3 bath. 
Our design has several attractive features: the quartz thermometer monitors the temperature of the He-3 which is
inherently in good thermal contact with the sample under study, the temperature reading is independent of the magnetic 
field, and the temperature is monitored in real time.
We provide evidence that the electronic temperature follows that of the He-3 bath, as measured by the quartz tuning fork. 

\section{The Helium-3 Immersion Cell}

Heatsinking of the leads in traditional setups of sample-in-vacuum
is done by wrapping copper wires with a thin enamel insulation around copper posts attached to different stages of the refrigerator, including the mixing chamber.
The limiting factor in cooling of the sample is therefore the Kapitza thermal resistance due to phonon mismatch between the copper post, wire insulation, epoxy used, and the wire itself. The Kapitza resistance is inversely proportional to the geometric area of overlap \cite{pobell}. Since wrapping of the wires permits a relatively modest overlap, one often finds a poor electron thermalization at milliKelvin temperatures.

\begin{figure}[t]
 \includegraphics[width=1\columnwidth]{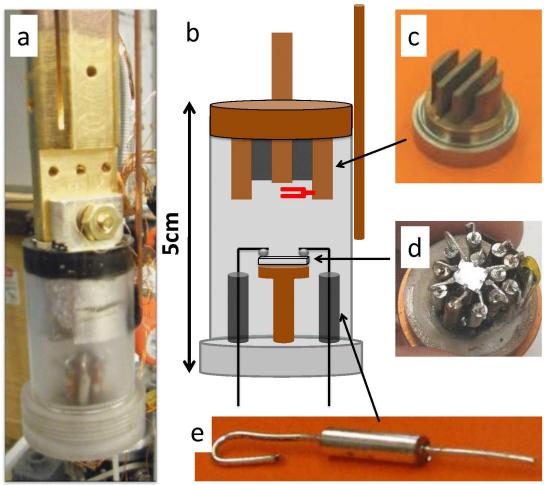}
 \caption{\label{f1}
Photo of the He-3 immersion cell (panel a). In the schematic drawing (panel b) copper is brown, silver is black, 
polycarbonate is translucent, and the quartz tuning fork is red.
The main heat exchanger prior to packing the silver powder in 
(panel c), a two-dimensional electron sample in a GaAs host soldered onto the heatsinks (panel d), and
an individual heatsink (panel e) are also shown.
}
\end{figure}

The use of sintered metals is a known way to increase surface area and therefore reduce the Kapitza resistance \cite{pobell,xia2000}. To cool our sample we use heatsinks made of 100-500nm silver powder sintered onto
0.7mm diameter silver wires \cite{xia2000}. One such heatsink is shown in Fig.1e and the sample attached onto several heatsinks can be seen in Fig.1d.
In order to take advantage of the increased surface area of the heatsinks we immerse them into a cryogenic liquid which assures thermal conduction as well as electrical insulation. While both liquid He-3 and superfluid He-4 provide excellent thermal conductivity at milliKelvin temperatures, we have chosen He-3 for two reasons. First, the viscosity of liquid He-3 is a strong function of the temperature and therefore it enables a viscometry based temperature measurement. Second, He-3 stays normal for the temperature range we access and therefore minimizes the chance of developing a superfluid leak from the cell into the vacuum space of the refrigerator. 

The immersion cell is shown in Fig.1.
It consists of four parts: a polycarbonate body with a thread on the bottom, a bottom cap (d) and a screw
fastening it to the body made of the same material, the sample heatsinks (e), and the main sinter heatexchanger (c). The cap which seals the cell from the bottom has to be removable to allow access to the sample. The seal is made by 
wetting the carefully machined polycarbonate body and cap with a thin layer of vacuum grease. 
The grease is seen in Fig.1d as a shiny circular ring on the rim of the bottom cap.
We have cooled down this cell more than 25 times, and we have not seen any evidence of He-3 leaking out into the vacuum. The main heat exchanger plugs the cell from the top and assures the cooling of the He-3. It consists of a piece of copper shown on Fig.1(c), with silver powder packed into it. The copper piece is sanded and electroplated with a thin layer of about 2 $\mu$m of silver in order to facilitate the adhesion of the silver powder. The powder is compressed to about 50\% of its original volume. We estimate a specific surface area of the sinters to be about 1 m$^2$/g. Finally there are a dozen sample heatsinks epoxied through the detachable bottom cap onto which the sample is soldered with indium (see Fig.1d). During soldering we use a copper alligator clip to prevent overheating of the heatsink sinters.
The cell is attached to the mixing chamber via an annealed high purity copper tail and filled with liquid He-3 through a capillary which is silver brazed onto copper posts attached to each cooling stage of the refrigerator.

\section{Thermometry based on quartz tuning fork viscometry}

At low milliKelvin temperatures the widely used Ruthenium Oxide (RuO) resistive thermometers are not
suitable for temperature measurement because of the loss of thermal contact and a strong dependence on the magnetic field applied \cite{nodar2010}. For thermometry down to the 5mK base temperature of our dilution refrigerator
we choose to measure the strongly temperature dependent viscosity of the He-3 liquid
that, together with our sample to be measured, is enclosed in the immersion cell. 
According to the Fermi liquid theory the viscosity of He-3 obeys the simple equation  $\eta T^2$=constant in the strongly degenerate limit, i.e. below about 30mK \cite{abrikosov1959}. It was found that a correction of the form
$\eta T^2(1-2.92T)$=constant can be used for temperatures up to 125mK \cite{bertinat1972}.

We measure the viscosity of He-3 with an oscillating quartz tuning fork which is fastened
to the main heatexchanger of our immersion cell. Quartz tuning fork thermometers have
recently been used in studies of quantum liquids \cite{blaauwgeers2000,clubb2004,pentti2009}  
and have a number of advantages. In our setup the quartz viscometer is in close vicinity of the sample
and measures the in-situ temperature of the He-3 bath in real time. We will show later that the electron temperature
follows that of the He-3 bath to the lowest temperatures.
Quartz viscometers are simpler to build than the previously used vibrating wire viscometers \cite{pobell}. In addition, quartz viscometers are easier to use than He-3 melting curve thermometers, are immune to radiofrequency heating, and dissipate very little heat. Since in vacuum the resonance of the quartz changes insignificantly with an applied magnetic field  \cite{rychen2000,clubb2004}, the strongly damped motion in the viscous He-3 is expected to be independent of the magnetic field. 

\begin{figure}
 \includegraphics[width=1\columnwidth]{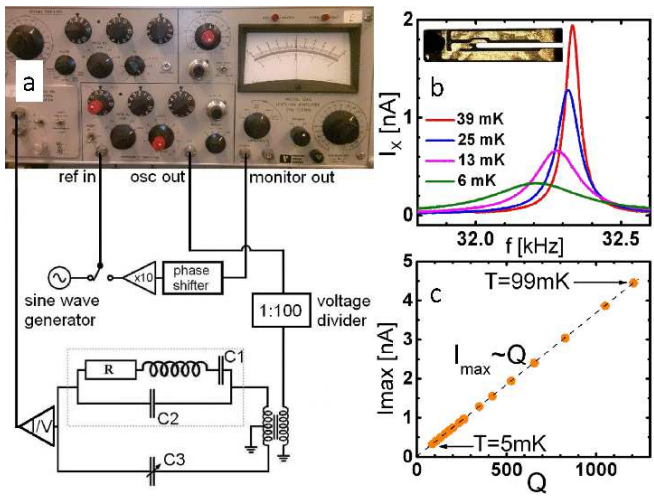}
 \caption{\label{f2}
The measuring circuit (panel a), the in-phase component of the resonant current I$_x$ as function of the
frequency of the driving oscillator at various temperatures (panel b), photo of the quartz tuning fork
(inset of panel b), and the amplitude of the current at the maximum of the resonance versus the
quality factor $Q$ (panel c). We used 5mV excitation.
}
\end{figure}

The motion of the quartz tuning fork in pure He-3 is described by the Stokes hydrodynamic model \cite{blaauwgeers2000,clubb2004,pentti2009}
which has recently been verified to be valid in the 5-100mK range by a cross-calibration against
He-3 melting curve thermometers \cite{blaauwgeers2000,pentti2009}. 
According to this model the quality factor of the oscillator $Q$ is inversely proportional
to the square root of the viscosity. Using the expression of the viscosity of He-3
below 125mK \cite{bertinat1972} we find $Q \propto T\sqrt{1-2.92T}$ . 
The proportionality constant is determined from a calibration against a commercial RuO thermometer at 40mK 
\cite{lakeshore}.

Figure 2a shows our measuring circuit for the quartz tuning fork viscometer. As seen in the dotted box in Fig.2a, the equivalent circuit of the quartz consists of an RLC series resonant circuit connected in parallel to a parasitic capacitance C2 \cite{rychen2000}. In order to measure the resonant part of the current we used a bridge configuration \cite{grober2000,saitoh2008} with a transformer \cite{transf} which effectively cancels the current due to the parasitic capacitance C2 of the quartz. This is achieved by tuning the capacitance C3 of a short semirigid coax to be equal to C2. The current is measured with a wide bandwidth preamplifier with a gain of $10^6$V/A \cite{femto} which is fed into a PAR model 124A lockin amplifier. The electrical cables of quartz are shielded using thin wall stainless steel tubing 
which is heatsunk to each stage of our refrigerator.

The circuit shown has two operating modes. When driving the reference channel of the lockin with an external generator we measure an in-phase resonant current with a perfect Lorentzian shape at any value of the temperature. Results of the frequency scan are shown in Fig.2b.
While this method of scanning the frequency through the resonance is a good way to measure $Q$, it is somewhat cumbersome and time consuming. Inspired by a circuit used to drive torsional oscillators in studies of superfluid 
films \cite{agnolet1989}, we employed a second operating mode which avoids the frequency scan and 
which utilizes a self-locking technique instead. This is achieved by switching the lockin frequency reference 
to the signal monitor output of the lockin. 
Using this technique we ensure that the frequency of the lockin automatically tracks the resonance frequency of the quartz. 
The phase shifter \cite{phase} is necessary to keep the frequency locked to the maximum of the Lorentzian curves rather than a different frequency close to it. We find that the current $I$ measured with the feedback loop containing the phase
shifter is equal to the maximum of the Lorentzian current versus frequency scans and therefore
the two diferent modes of running the quartz yield the same current output and hence the same temperature.
In order to convert current into temperature, we first relate the measurened current $I$
to $Q$. As seen in Fig.2c, this functional dependene is measured to be linear and therefore the current at resonance is 
$I \propto Q \propto T \sqrt{1-2.92T}$. This technique substantially simplifies the measurement procedure and makes continuous monitoring of the temperature possible. We note that we did not succeed in running our quartz in the self-locked mode using an SRS lockin.

\section{Characterizing the Electromagnetic Environmnet of the Sample}

\begin{figure}[t]
 \includegraphics[width=.8\columnwidth]{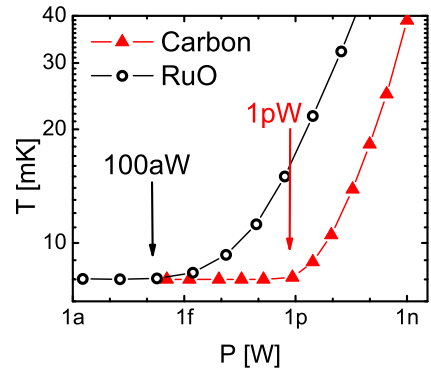}
 \caption{Self-heating curves of a RuO and carbon composition resistors immersed into He-3
show the minimum power levels causing measurable self-heating at 8mK are 100aW and 1pW, respectively. 
 \label{Fig3}}
 \end{figure} 
 
As discussed in the introduction, microwave heating is often a source of saturation of the electronic temperature
in the milliKelvin range. In order to filter these waves we use a commercial room temperature filter \cite{filter}
mounted in an aluminum housing, equipped with a Fischer connector which is directly plugged in to its mating
connector on top of the refrigerator. Electrical connection from the top of the refrigerator to the mixing chamber is
via constantan wires and from the mixing chamber to the immersion cell using polyimide coated copper wires.
A second cold radiofrequency filter is made by enclosing these copper wires in silver epoxy. The same epoxy fastens
the wires onto the copper tail connecting the immersion cell to the mixing chamber.

To characterize radiofrequency and possible ground loop heating in our setup 
we use a resistive thermometer soldered onto the heatsinks which are immersed into He-3.
This is useful since cooling electrons in resistive thermometers encounters the same problems as cooling electrons in nanostructures \cite{nodar2010}.
We considered two thermometers: a carbon-based thermometer described in Ref.\cite{nodar2010}
and a RuO thermometer \cite{ruo}. The carbon resistor was thinned down by removing its phenolic 
protection for short response times. As discussed in Ref.\cite{nodar2010}, such a carbon thermometer has
an excellent thermal contact to its environment. In order to compare the thermal contact of the carbon and RuO
thermometers we measured their self-heating curves at a constant cell temperature of 8mK as indicated by the quartz
thermometer. These self-heating curves, shown in Fig.3, are performed by applying increasingly larger 
excitation currents and
converting the measured resistances to temperatures using the calibration done in the limit of no self-heating. Since the RuO thermometer starts self-heating at the power of 100aW, we conclude that it is more suitable to estimate
the residual heating and that the above value is an upper bound for the spurious radiofrequency power in our setup.

\section{Testing of the Immersion Cell} 

\begin{figure}[b]
 \includegraphics[width=.8\columnwidth]{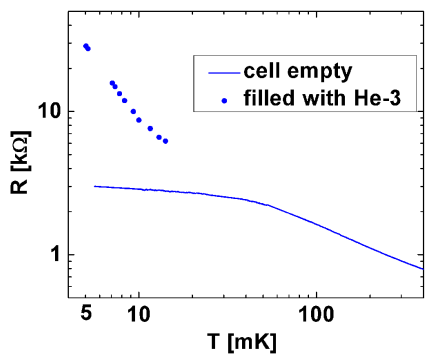}
 \caption{Filling of the immersion cell with liquid He-3 causes a large change in the resistance of
 a carbon thermometer soldered onto the heatsinks from 3k$\Omega$ to 30k$\Omega$ at the mixing chamber
 temperature of 5mK.
 \label{Fig4}}
 \end{figure} 
 
In order to assess the effectiveness of our immersion cell
we measured the resistance of the carbon thermometer mounted onto the heatsinks as a function 
of mixing chamber temperature in two different configurations.
First, we measured the resistance when the cell was intentionally left empty. In this configuration the carbon resistor is cooled only through the copper electrical leads. This configuration mimicks a traditional setup of sample-in-vacuum. 
As shown in Fig.4, we found that the resistance of the carbon thermometer saturates at 3k$\Omega$  below about 30mK.
We conclude that in this configuration we are unable to cool our thermometer below 30mK even though our 
mixing chamber was at 5mK.
 
In the second configuration we filled the immersion cell with liquid He-3. We observed a dramatic change in the resistance of the thermometer from 3k$\Omega$ to 30k$\Omega$ at 5mK mixing chamber temperature.
We interpret this tenfold increase of the resistance as evidence for a much reduced electron temperature 
in the thermometer when the cell is filled with He-3.

\begin{figure}[b]
 \includegraphics[width=1\columnwidth]{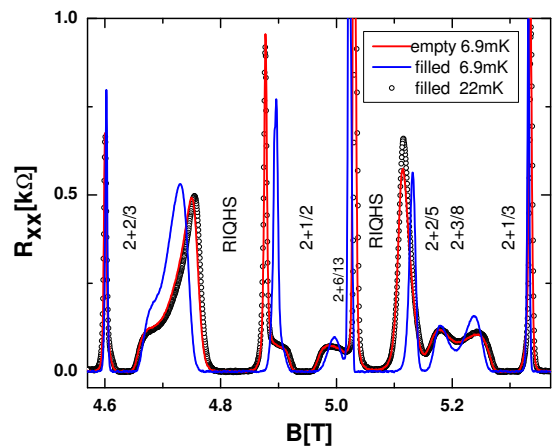}
 \caption{
Filling of the immersion cell with liquid He-3 while keeping the mixing chamber at 6.9mK
results in a large change in the magnetoresistance of a two-dimensional electron gas. 
We marked the various fractional quantum Hall and the reentrant integer quantum Hall states (RIQHS). 
Also shown is the good overlap of the traces taken at 22mK with the cell filled and 
at 6.9mK fridge temperature with the cell empty. The inset indicates the state of the cell and the temperature 
of the mixing chamber.
  \label{Fig5}}
 \end{figure}

For further testing we performed a similar set of measurements on a two-dimensional electron gas in a
GaAs quantum well mounted on the heatsinks. The size of the sample 
is $4\times4\times0.5$mm$^3$, its electron density is $3.0\times10^{11}$cm$^{-2}$ and mobility $32\times10^6$cm$^2$/Vs. 
We compared the magnetroresistance R$_{xx}$ of this sample with the refrigerator cooled to 6.9mK but the immersion
cell empty to that measured with the cell filled with He-3 held at various temperatures.
As seen in Fig.5, the magnetoresistance trace with the cell filled is much 
improved as compared to the trace taken with the cell empty while keeping the fridge at 6.9mK. 
For example the width of the R$_{xx}$ minima for the reentrant integer quantum Hall states
is wider and the depth of R$_{xx}$ of the fractional quantum Hall state at Landau level filling factor 
$\nu=2+2/5$ and $2+6/13$ is enhanced when the cell is filled.
Furthermore, the trace with the cell empty has a very good overlap with the
trace with the cell filled with He-3 and held at 22mK. We conclude that when the sample is cooled only through the 
measurement wires, its effective electron temperature is 22mK even when the refrigerator is very cold and that
filling the cell with He-3 results in a lower electron temperature.

In order to directly compare the temperature of the electrons to that of the He-3 bath as measured by the quartz viscometer, we measure the dependence of the magnetoresistance of the
$\nu=2+3/8$ developing fractional quantum Hall state as function of the temperature as measured by the quartz viscometer.
It is well known that if R$_{xx}$ is not vanishingly small, i.e. outside of the region of variable range hopping,
the magnetoresistance has an activated temperature dependence of the form $R_{xx} \propto \exp(-T/2\Delta)$
\cite{tsui}. As seen in Fig.6a,
the magnetoresistance of the $\nu=2+3/8$ fractional quantum Hall state follows such a functional dependence.
Furthermore, as seen in Fig.6b, the width of the $\nu=2+1/2$ plateau does not saturate to
temperatures as low as 4mK reached by demagnetizing our copper tail from the starting field of 10T.
We thus conclude that the electron temperature follows that of the He-3 bath as measured by the quartz viscometer
to as low as 4mK. The use of this setup has recently enabled the observation of a new fractional quantum Hall state
in the second Landau level at $\nu=2+6/13$ \cite{kumar2010}.

\begin{figure}
 \includegraphics[width=.9\columnwidth]{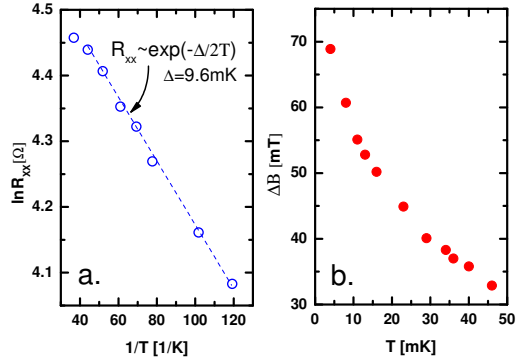}
 \caption{
The activated temperature dependence of the magnetoresistance at $\nu=2+3/8$ (panel a)
and the monotonic width of the $\nu=2+1/2$ plateau as function of the temperature (panel b)
provide evidence that the electron temperature follows that of the He-3 bath. 
\label{Fig6}}
 \end{figure}
 
\section{Conclusions}

We have built a He-3 immersion cell which allows cooling of the electrons in semiconductor nanodevices to temperatures of a few milliKelvin. Such a low bath temperature is measured using a quartz He-3 viscometer. We have tested the performance of the cell subjected to a magnetic field by measuring the fractional quantum Hall effect. The activation measurement of the $\nu=2+3/8$ fractional quantum Hall state provides evidence that the temperature of the charge carriers follows that of the He-3 bath, as measured by the quartz viscometer.

\begin{acknowledgments}
We thank J.S. Xia and N.S. Sullivan at the microKelvin Facility of the National High Magnetic Field Laboratory in Gainesville, Florida for annealing the copper tail used to connect our experimental cell to the mixing chamber. 
G.A.C. and N.S. were supported on NSF DMR-0907172 and 
M.J.M. acknowledges the support of the Miller Family Foundation. 
\end{acknowledgments}


\begin{thebibliography}{l}
\bibitem{stoner1964} E.C. Stoner, Rep. Prog. Phys. \textbf{11}, 43 (1947).
\bibitem{candido2004} L. C\^andido, B. Bernu, and D.M. Ceperley, Phys. Rev. B \textbf{70}, 094413 (2004).
\bibitem{moore91} G. Moore and N. Read, Nucl. Phys. B \textbf{360}, 362-396 (1991).
\bibitem{pan02} W. Pan, H.L. Stormer, D.C. Tsui, L.N. Pfeiffer, K.W. Baldwin, and K.W. West, Phys. Rev. Lett. \textbf{88}, 176802 (2002).
\bibitem{eisen02} J.P. Eisenstein, K.B. Cooper, L.N. Pfeiffer, and K.W. West, Phys. Rev. Lett. \textbf{88}, 076801 (2002).
\bibitem{xia04} J.S. Xia, W. Pan,  C.L. Vicente, E.D. Adams, N.S. Sullivan, H.L. Stormer, D.C. Tsui, L.N. Pfeiffer, K.W. Baldwin, and K.W. West, Phys. Rev. Lett. \textbf{93}, 176809 (2004).
\bibitem{kumar2010} A. Kumar, G.A. Cs\' athy, M.J. Manfra, L.N. Pfeiffer, and K.W. West,
Phys. Rev. Lett. \textbf{105}, 246808 (2010).
\bibitem{willett09} R.L. Willett, L.N. Pfeiffer, and K.W. West, Proc. Natl. Acad. Sci. U.S.A., \textbf{106}, 8854 (2009).
\bibitem{stern10} A. Stern, Nature \textbf{464}, 187-193 (2010), and references therein.
\bibitem{pobell}	see for example in F. Pobell, "Matter and methods at low temperatures", 3rd ed., Springer 2007.
\bibitem{wennberg1986} A.K.M. Wennberg, S.N. Ytterboe, C.M. Gould, H.M. Bozler, J. Klem, H. Morkoc, Phys. Rev. B \textbf{34}, 4409 (1986).
\bibitem{xia2000}	J.S. Xia, E.D. Adams, V. Shvarts, W. Pan, H.L. Stormer, and D.C. Tsui, Physica B \textbf{280}, 491  (2000); W. Pan, J.S. Xia, V. Shvarts, D.E. Adams, H.L. Stormer, D.C. Tsui, L.N. Pfeiffer, K.W. Baldwin, and K.W. West, Phys. Rev. Lett. \textbf{83}, 3530 (1999).
\bibitem{nodar2010}	N. Samkharadze, A. Kumar, G.A. Cs\' athy, J. Low Temp. Phys. \textbf{160}, 246 (2010).
\bibitem{abrikosov1959}	A.A. Abrikosov, I.M. Khalatnikov, Rep. Prog. Phys. \textbf{22}, 329 (1959);
T.A. Alvesalo, H.K. Collan, M.T. Loponen, O.V. Lounasmaa, M.C. Veuro, J. Low Temp. Phys. \textbf{19}, 1 (1975);
D.A. Ritchie, J. Saunders, and D.F. Brewer, Phys. Rev. Lett. \textbf{59}, 465 (1987).
\bibitem{bertinat1972}	M.P. Bertinat, D.S. Betts, D.F. Brewer, G.J. Butterworth,  J. Low Temp. Phys. \textbf{16}, 479 (1974).
\bibitem{blaauwgeers2000}	R. Blaauwgeers, M. Blazkova, M. Clovecko, V.B. Eltsov, R. de Graaf, J. Hosio, M. Krusius, D. Schmoranzer, W. Schoepe, L. Skrbek, P. Skyba, R.E. Solntsev, and D.E. Zmeev, J. Low. Temp. Phys. \textbf{146}, 537 (2000).
\bibitem{clubb2004}	D. Clubb, O.V.L. Buu, R.M. Bowley, R. Nyman, J.R. Owers-Bradley, J. Low Temp. Phys. \textbf{136}, 1 (2004).
\bibitem{pentti2009} E. Pentti, J. Rysti, A. Salmela, A. Sebedash, and J. Tuoriniemi, Helsinki University of Technology,
Low Temperature Laboratory Publications, Report TKK-KYL-021 (2009).
\bibitem{rychen2000}	J. Rychen, T. Ihn, P. Studerus, A. Herrmann, K. Ensslin, H.J. Hug, P.J.A. van Schendel, and H.J. Guntherodt, Rev. Sc. Instr. \textbf{71}, 1695 (2000).
\bibitem{lakeshore}	model RX-202A, Lakeshore  Cryotronics, www.lakeshore.com
\bibitem{grober2000}	R.D Grober, J. Acimovic, J. Schuck, D. Hessman, P.J. Kindlemann, J. Hespanha, A.S. Morse, 
Rev.Sc. Instr. {\textbf 71}, 2776 (2000).
\bibitem{saitoh2008} K. Saitoh, K. Hayashi, Y. Shibayama, and K. Shirahama, J. Low Temp. Phys. \textbf{150}, 561 (2008).
\bibitem{transf} Triad model SP-67, purchased from www.digikey.com
\bibitem{femto}	Femto Messtechnik GmbH, Current Preamplifier Model DLCPA-200.
\bibitem{agnolet1989} G. Agnolet, D.F. McQueeney, and J.D. Reppy, Phys. Rev. B \textbf{39}, 8934 (1989).
\bibitem{phase}	J.S. Lopez, A.A. Melo, V.S. Oliveira, Phys. Educ. \textbf{17}, 238 (1982).
\bibitem{filter} Spectrum Control, part number 56-721-012.
\bibitem{ruo} KOA 649$\Omega$, part number 660-RK73H2ELTD6490F from www.mouser.com
\bibitem{tsui} D.C. Tsui, H.L Stormer, and A.C. Gossard, Phys. Rev. B \textbf{25}, 1405 (1982);
G.S. Boebinger, A.M. Chang, H.L. Stormer, and D.C. Tsui, Phys. Rev. Let.. \textbf{55}, 1606 (1985).





\end{thebibliography}

\end{document}